\documentstyle[mnras_cite,psfig,onecolumn]{mn}
\def\d{{\rm d}}
\begin{document}

\title{Measuring the Distribution of Galaxies Between Halos} 
\author[Andrew~J.~Benson]{Andrew~J.~Benson$^1$ \\
1. California Institute of Technology, MC 105-24, Pasadena, CA 91125, U.S.A. (e-mail: abenson@astro.caltech.edu)}

\maketitle

\begin{abstract}
We develop a method to measure the probability, $P(N;M)$, of finding
$N$ galaxies in a dark-matter halo of mass $M$ from the theoretically
determined clustering properties of dark-matter halos and the
observationally measured clustering properties of galaxies. Knowledge
of this function and the distribution of the dark matter completely
specifies all clustering properties of galaxies on scales larger than
the size of dark-matter halos. Furthermore, $P(N;M)$ provides strong
constraints on models of galaxy formation, since it depends upon the
merger history of dark-matter halos and the galaxy-galaxy merger rate
within halos. We show that measurements from a combination of the
2MASS and SDSS or 2dFGRS datasets will allow $P(N;M)$ averaged over
halos occupied by bright galaxies to be accurately measured for
$N=0$--2.
\end{abstract}

\section{Introduction}

Recent work on the clustering properties of galaxies has focussed on
the connection between galaxies and dark matter halos, using
theoretical models of galaxy formation
\cite{kns,kauffdummy,diaferio99,ajbdummy,somerville00,seljak00} or
observational data \cite{peacock00} to determine the number of
galaxies that reside within halos of given mass. Models of this type
have been successful in explaining the near power-law nature of the
galaxy-galaxy correlation function \cite{kauff99a,benson00}, and the
strong clustering of Lyman-break galaxies at $z\approx 3$
\cite{fabio98,baugh98,wechsler00}.

In this context, \scite{benson00} calculated the quantity $P(N;M)$,
the probability of finding $N$ galaxies brighter than a specified
luminosity, $L_0$, in a dark matter halo of mass $M$, from the
galaxy-formation model of \scite{cole00}. This quantity is
particularly powerful since, once a model for the distribution of
dark-matter halos is chosen, $P(N;M)$ fully determines all clustering
properties of galaxies on scales larger than the size of dark matter
halos (on smaller scales the spatial distribution of galaxies within
individual halos becomes important). As such, $P(N;M)$ may be thought
of as a complete description of the galaxy---dark-matter bias
including any non-linearity and stochasticity. Furthermore, if
$P(N;M)$ can be measured from the observed clustering pattern of
galaxies it provides a direct and powerful constraint for models of
galaxy formation since it is sensitive to the merger history of
dark-matter halos, and to the rate of galaxy-galaxy mergers within
halos.

In this paper we describe how $P(N;M)$ may be measured directly from a
volume-limited galaxy-redshift survey by using a counts-in-cells
analysis to determine the probability of finding $N$ galaxies in a
cell, $S(N)$. The remainder of this paper is laid out as follows. In
\S\ref{sec:method} we describe our method and give the formulae
relating $P(N;M)$ and $S(N)$ for all $N$. In \S\ref{sec:mocks} we
investigate how well $P(N;M)$ can be measured from a combination of
the Two Micron All Sky Survey (\pcite{skrutskie95}; 2MASS) and
Two-degree Field Galaxy Redshift Survey (\pcite{dalton00}; 2dFGRS) or
Sloan Digital Sky Survey (\pcite{blanton00}; SDSS) datasets using the
mock galaxy catalogues of \scite{benson00}, and finally in
\S\ref{sec:discuss} we present our conclusions.

\section{Method}
\label{sec:method}

We will assume that the galaxy population of a dark matter halo is
determined only by the mass of that halo. Whilst it is the
distribution of halo masses which varies most significantly as a
function of environment \cite{lemson99} other quantities are also
known to correlate with environment, for example the concentration of
the halo \cite{bullock99}. In practice the properties of galaxies may
depend upon such variables thereby altering the clustering properties
of the galaxies. In principle, other variables could be included in
our analysis by defining a function $P(N;M,{\bf x})$, where ${\bf x}$
represents other variables upon which the properties of galaxies may
depend. However, current datasets are insufficient to allow meaningful
measurements of such a function to be made and so we will restrict
ourselves to considering $P(N;M)$ only at present.

We must also assume that $P(0,M)=1$ for all $M<M_0$; i.e. halos below
mass $M_0$ never contain any galaxies brighter than $L_0$. This is a
reasonable and necessary assumption --- if halos of arbitrarily low
mass could host bright galaxies then, since there are an infinite
number of halos per unit volume (at least according to the
Press-Schechter theory) there would be an infinite number of galaxies
per unit volume. Having made this assumption we can ignore halos of
mass less than $M_0$ as they make no contribution to the galaxy
population that we are considering.

How may we determine $M_0$ for a given galaxy population? One approach
would be to make use of dynamical mass estimates
(e.g. \pcite{vogt97}). However, these are not available for all types
of galaxy. An alternative method is to use galaxy samples selected at
near-infrared wavelengths from which we can infer a stellar mass from
the sample magnitude limit \cite{kc98}. Then
\begin{equation}
M_0\geq {\Omega_0 \over \Omega_{\rm b}} M_\ast,
\label{eq:mzero}
\end{equation}
where $M_\ast$ is the stellar mass. This lower limit on $M_0$
corresponds to the case where the entire gaseous mass of a halo is
turned into stars. The halo must have at least this mass to make the
observed galaxy. The conversion from K-band light to stellar mass is
uncertain by a factor of approximately two \cite{jarle00}, so $M_0$
should realistically be taken to be two to three times lower than the
value inferred from eqn.~(\ref{eq:mzero}).

While any clustering statistic can be written in terms of $P(N;M)$ and
the clustering properties of dark matter halos (for example, the
two-point correlation function expressed in terms of $P(N;M)$ is given
in Appendix~\ref{ap:xi}) a particularly simple relation can be found
for $S(N)$, the probability of finding $N$ galaxies brighter than
$L_0$ in a cell of given size and shape.  While these statistics can
in principle reveal $P(N;M)$ for any $N$ and for a range of $M$, in
practice measurement is severely limited by unavoidable noise in the
data as will be shown in \S\ref{sec:mocks}. Nevertheless, useful
constraints can still be obtained from this analysis. In the remainder
of this section we develop the relations necessary to determine
$P(N;M)$ for all $N$ and $M$, but will only make use of the simplest
forms of these relations in \S\ref{sec:mocks}.

The probability of finding $N$ galaxies in a cell of given size and
geometry can be expressed in terms of the probability of finding a
certain combination of halos in that cell and the probabilities of
finding different numbers of galaxies in each of those halos. In order
to measure $P(N;M)$ it is necessary to divide halos into a number of
mass ranges, or bins. We will then refer to the mean value of $P(N;M)$
averaged over all halos in mass bin $i$ as $P_i(N)$, such that
$P_i(N)$ is the probability of finding $N$ galaxies in a halo selected
at random from mass bin $i$, i.e.
\begin{equation}
P_i(N) = {\int_{M_i}^{M_{i+1}} P(N;M) {\d n \over \d M}(M) \d M \over \int_{M_i}^{M_{i+1}} {\d n \over \d M}(M) \d M},
\end{equation}
where $M_i$ is the lower bound of the $i^{\rm th}$ mass bin.

Let $S(N)$ be the probability of finding $N$ galaxies in a cell (of
given size and geometry). For a particular choice of cosmology and
dark matter let $Q(N_1,N_2,\ldots,N_n)$ be the probability of finding
the centres of $N_1$ halos in mass bin 1, $N_2$ in bin 2 etc. in a
cell, where we have used a total of $n$ mass bins. (We take the centre
of mass to define the halo centre.) Note that in general
$Q(N_1,N_2,\ldots,N_n) \ne Q(N_1)Q(N_2)\ldots Q(N_n)$ since the
distribution of halos is typically correlated. Note that $S(N)$ is an
observationally measurable quantity, and $Q(N_1,N_2,\ldots,N_n)$ can
be obtained from a structure formation model. As we show below, these
two quantities are related, and that relation depends upon
$P(N;M)$. Measurement of $S(N)$ therefore allows us to measure
$P(N;M)$.

We can write $S(N)$ as the sum over all possible combinations of
$N_1,N_2,\ldots ,N_n$ of $Q(N_1,N_2,\ldots,N_n)$ multiplied by the
probability of finding $i_1$ galaxies in the first halo, $i_2$ in the
second etc. summed over all combinations of $i_1,i_2,\ldots$ which
satisfy the constraint $\sum_j i_j=N$ (i.e. only those combinations
which produce the correct number of galaxies in the cell contribute to
the total probability).

For example, $S(0)$ is given by
\begin{equation}
S(0) = \sum_{N_1=0}^\infty \sum_{N_2=0}^\infty \ldots \sum_{N_n=0}^\infty Q(N_1,N_2,\ldots,N_n) \prod_{j=1}^n P_j^{N_j}(0),
\end{equation}
while $S(1)$ and $S(2)$ are given by
\begin{equation}
S(1) = \sum_{N_1=0}^\infty \sum_{N_2=0}^\infty \ldots \sum_{N_n=0}^\infty Q(N_1,N_2,\ldots,N_n) \sum_{i_1=1}^n C(N_1^{(0)},N_1^{(1)},N_2^{(0)},N_2^{(1)}) {P_{i_1}(1) \over P_{i_1}(0)} \prod_{j=1}^n P_j^{N_j}(0)
\end{equation}
and
\begin{eqnarray}
S(2) & = & \sum_{N_1=0}^\infty \sum_{N_2=0}^\infty \ldots \sum_{N_n=0}^\infty Q(N_1,N_2,\ldots,N_n) \left( \sum_{i_1=1}^n C(N_1^{(0)},\ldots ,N_1^{(2)},N_2^{(0)},\ldots ,N_2^{(2)}) {P_{i_1}(2) \over P_{i_1}(0)} \right. \nonumber \\
 & & \left. + \sum_{i_1=1}^n \sum_{i_2=1}^n C(N_1^{(0)},N_1^{(1)},N_2^{(0)},N_2^{(1)}) {P_{i_1}(1) \over P_{i_1}(0)}{P_{i_2}(1) \over P_{i_2}(0)} \right) \prod_{j=1}^n P_j^{N_j}(0),
\end{eqnarray}
where $N_i^{(j)}$ is the number of times a halo in mass bin $i$ is
populated by $j$ galaxies and $C(N_{i_1}^{(0)},N_{i_1}^{(1)}\ldots)$
is the number of distinct permutations of each term which contribute
to the probability. The weighting factor
$C(N_{i_1}^{(0)},N_{i_1}^{(1)}\ldots)$ is the number of ways to
populate the available halos with the galaxies divided by the number
of times such terms appear in the summation. In general,
\begin{equation}
C(N_{i_1}^{(0)},\ldots ,N_{i_1}^{(k)},\ldots ,N_{i_n}^{(0)},\ldots ,N_{i_n}^{(k)}) = \left. \prod_{l=1}^n {N_{i_l}! \over \prod_{j=0}^k N_{i_l}^{(j)}!} \right/ \prod_{m=1}^k {\left(\sum_{l=1}^n N_{i_l}^{(m)}\right)! \over \prod_{l=1}^n N_{i_l}^{(m)}!} = \prod_{l=1}^n {N_{i_l}! \over  N_{i_l}^{(0)}!\left(\sum_{m=1}^n N_{i_m}^{(l)}\right)!},
\end{equation}
where $N_i=\sum_{j=0}^\infty N_i^{(j)}$. For example,
\begin{equation}
C(N_{i_1}^{(0)},N_{i_1}^{(1)})={\left(N_{i_1}^{(0)}+N_{i_1}^{(1)}\right)! \over N_{i_1}^{(0)}! N_{i_1}^{(1)}!}.
\end{equation}

As expected, $S(N)$ depends only upon those $P_i(j)$ for which $j\leq
N$. Therefore, we may begin by finding the $P_i(0)$'s using the
expression for $S(0)$, then proceed to find the $P_i(1)$'s using the
expression for $S(1)$ and the previously calculated $P_i(0)$ and so
on. Each expression therefore involves $n$ unknowns (for $S(N)$ these
are the $P_i(N)$), and so we must have a measure of $S(N)$ for at
least $n$ different cell sizes to solve the equations. While the
above equations cannot be solved analytically for the $P_i(N)$,
solutions can be found relatively simply using Powell's method
\cite{numrec} to minimize the quantity $\chi^2=\sum_{i} ([S_i^{\rm
(obs)}(N)-S_i^{\rm (model)}(N)]/\Delta S_i^{\rm obs}(N))^2$ for
example, where the sum is taken over all cell sizes considered.

In general, the expression for $S(N)$ will be of the form
\begin{eqnarray}
S(N) & = & \sum_{N_1=0}^\infty \sum_{N_2=0}^\infty \ldots \sum_{N_n=0}^\infty Q(N_1,N_2,\ldots,N_n) \Big( \sum_{{i_1}=1}^n C(N_1^{(0)},\ldots ,N_1^{(N)}) {P_{i_1}(N) \over P_{i_1}(0)}  \nonumber \\
 & & +  \sum_{{i_1}=1}^n \sum_{{i_2}=1}^n \left[ C(N_1^{(0)},\ldots ,N_1^{(N-1)},N_2^{(0)},\ldots ,N_2^{(N-1)}) {P_{i_1}(N-1) \over P_{i_1}(0)}{P_{i_2}(1) \over P_{i_2}(0)} \right.  \nonumber \\
 & & \left. + C(N_1^{(0)},\ldots ,N_1^{(N-2)},N_2^{(0)},\ldots ,N_2^{(N-2)}) {P_{i_1}(N-2) \over P_{i_1}(0)}{P_{i_2}(2) \over P_{i_2}(0)} \right] \nonumber \\
 & &  + \hbox{3 halo terms} + \hbox{4 halo terms} + \ldots + N\hbox{ halo terms} \Big) \prod_{j=1}^n P_j^{N_j}(0),
\label{eq:maineq}
\end{eqnarray}
where the expression ``$N$ halo terms'' refers to all terms
corresponding to galaxies shared between $N$ different halos (i.e. the
first two sums in the above expression are therefore ``1 halo terms''
and ``2 halo terms'').

At this point it is instructive to briefly consider the assumptions
made in obtaining the above relations. Firstly we have assumed that
all galaxies lie at the centre of the halo they occupy. Then, a halo
being in a cell guarantees that any galaxies it contains are also in
the cell. In reality galaxies are likely to be spread throughout the
halo with some unknown spatial distribution, and so some galaxies may
lie outside of the cell even though their halo centre is inside (and
conversely some galaxies may lie inside even though their halo centre
is outside). While our analysis could be extended to account for such
``edge effects'' this would require us to assume a distribution for
galaxies within individual halos. We prefer to concentrate on scales
where these effects are negligible. In \S\ref{sec:mocks} we
demonstrate that edge effects are an insignificant source of error.

Secondly we assume that the galaxy occupancy of all halos in a mass
bin is well described by a single set of $P_i(N)$. Providing $P(N;M)$
varies little across the mass bin this is a reasonable
assumption. However, as we will see in \S\ref{sec:mocks} noisy data
may limit us to considering a single mass bin, extending from $M_0$ to
infinity, for which the above assumption is unlikely to hold
true. While we implicitly assume that the $P_i(N)$ are independent of
the number of halos found in a cell the Press-Schechter \cite{PS74}
formalism tells us that high density regions of the Universe will
contain preferentially higher mass halos than low density
regions. Consequently cells which contain many halos will
preferentially contain high mass halos, while in cells containing few
halos the halos are likely to be of low mass. If, for example,
$P(0;M)$ is a decreasing function of $M$ then cells with few halos
(which are typically the most abundant) will contain zero galaxies
more often than our model assumes. The resulting increase in $S(0)$
can be seen in the synthetic datasets used in
\S\ref{sec:mocks}. While this has a non-negligible effect on $S(0)$,
particularly for large cell sizes, the value of $P(0)$ recovered is
quite insensitive to this since most of the signal comes from small
cell sizes.

\section{Application to Synthetic Datasets}
\label{sec:mocks}

Perhaps the most suitable dataset to apply this technique to will be a
combination of the 2MASS survey with a large redshift survey (e.g. the
2dFGRS or the SDSS). The 2MASS survey provides near-infrared
photometry which allows $M_0$ to be estimated, but must be
complemented by a redshift survey in order to provide a 3D map of the
galaxy distribution.\footnote{While $P(N;M)$ could be measured from a
2D dataset, the 3D information will provide a much stronger
constraint.} A volume limited 2MASS sample of galaxies brighter than
$M_{\rm K}-5\log h=-23.5$ would have a volume of order $3\times
10^6h^{-3}$Mpc$^3$ in the 2dFGRS survey area (or around four times
this volume in the SDSS survey area). As this is very similar to the
volume of the GIF $\Lambda$CDM N-body simulation used by
\scite{benson00} we will use their synthetic galaxy catalogues to
estimate how well $P(N;M)$ could be recovered from such a dataset. We
do not attempt here to reproduce the full details of the survey
geometry or selection function, but merely consider a synthetic
dataset with comparable volume and number density of galaxies in order
to estimate the accuracy with which $P(N;M)$ may be recovered from
such a survey.

We consider only galaxies brighter than $M_{\rm K}-5\log h=-23.5$ to
ensure that we need only consider halos which are well resolved by the
GIF simulation. These galaxies live in halos with masses greater than
$10^{12}h^{-1}M_\odot$ in this model (the particle mass in the GIF
$\Lambda$CDM simulation is $1.4\times
10^{10}h^{-1}M_\odot$). Inferring $M_0$ from the K-band magnitude of
the galaxies we find $M_0=8\times 10^{11}h^{-1}M_\odot$. We therefore
conservatively set $M_0=4\times 10^{11}h^{-1}M_\odot$. We consider
only one bin of halo mass, i.e. all halos more massive than $4\times
10^{11}h^{-1}M_\odot$. While this technique can in principle be
applied to several halo mass bins we find that in practice this is
very difficult. Typically the values of $P_i(N)$ for the more massive
bins are poorly constrained since there are very few halos in the mass
range, or else the solutions of eqn.~(\ref{eq:maineq}) for different
cell sizes are degenerate in the $P_i(N)$'s and so only allow certain
combinations of $P_i(N)$'s to be accurately measured. Very large
datasets, with correspondingly small errors may allow a measurement of
$P_i(N)$ in more than one mass bin, although this will probably
require a treatment of edge effects which must eventually become the
dominant source of error.

We measure $S(N)$ in cubic cells of side $L_{\rm cube}=5.0$, $7.5$,
$10.0$, $12.5$, $15.0$, $17.5$ and $20.0h^{-1}$Mpc, and measure $Q(N)$
for the same cell sizes. Both $S(N)$ and $Q(N)$ are calculated for
galaxy/halo positions in redshift space. For smaller cubes edge
effects begin to become a significant source of error, while for
larger cubes the GIF simulation contains very few independent volumes.

\begin{figure}
\begin{tabular}{cc}
\psfig{file=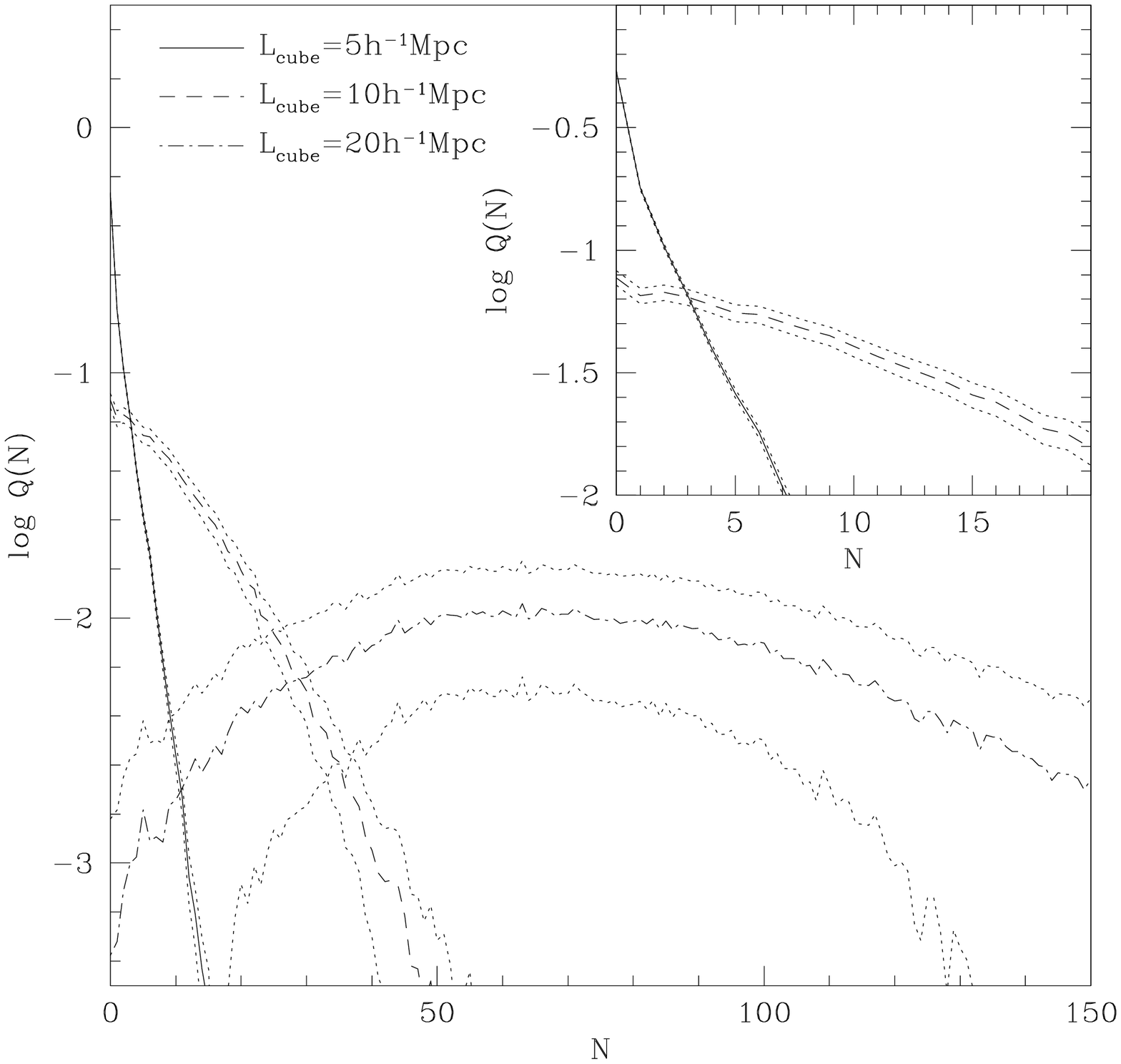,width=80mm} & \psfig{file=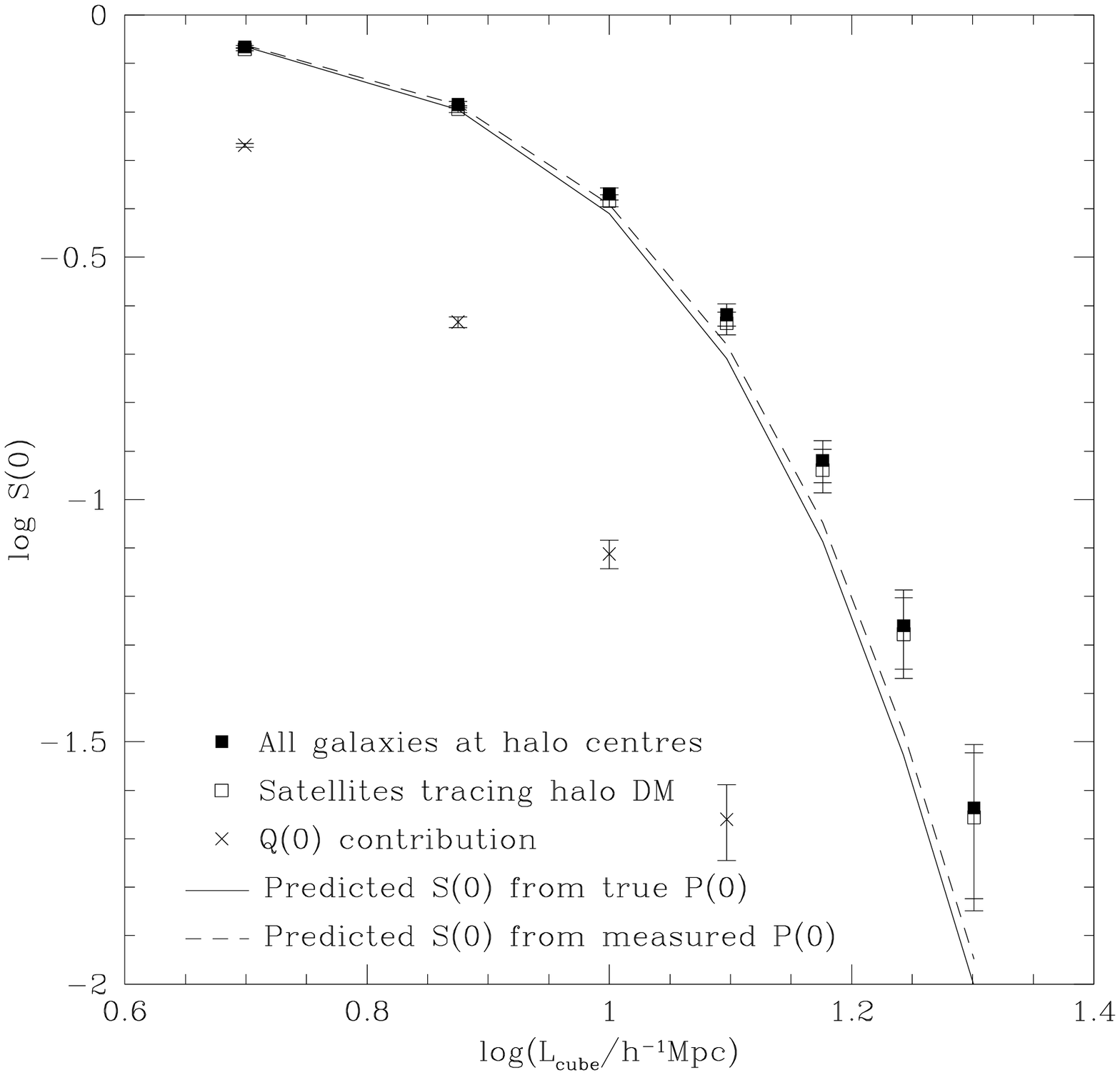,width=80mm}
\end{tabular}
\caption{\emph{Left-hand panel:} The probability of finding $N$ halos
more massive than $4\times 10^{11}h^{-1}M_\odot$ in cubes of sides 5
(solid line), 10 (dashed line) and $20h^{-1}$Mpc (dot-dashed line) in
redshift-space. Dotted lines indicate errors on these quantities
assuming Poisson statistics. The inset shows an expanded view of the
low-$N$ region. \emph{Right-hand panel:} The probability of finding
zero galaxies brighter than $M_{\rm K}-5\log h=-23.5$ in cubic cells
of side $L_{\rm cube}$ in redshift-space in the simulations of
\protect\scite{benson00}. Solid squares show the result when all
galaxies are placed at the centres of dark matter halos, while open
squares indicate the result when satellite galaxies are made to trace
the dark matter in their halo. Errors are calculated assuming Poisson
statistics. The solid line shows $S(0)$ calculated from the measured
$Q(N)$ (as shown in the left-hand panel) and the value of $P(0)$
measured directly from the models of \protect\scite{benson00}. Crosses
with errorbars show the contribution of $Q(0)$ to $S(0)$. The $Q(0)$
contribution is around 65\% for $L_{\rm cube}=5h^{-1}$Mpc, and falls
for larger values of $L_{\rm cube}$.}
\label{fig:sn}
\end{figure}

The left-hand panel of Fig.~\ref{fig:sn} shows $Q(N)$ for $L_{\rm
cube}=5$, 10 and $20h^{-1}$Mpc, while the right-hand panel shows
$S(0)$ (squares) and $Q(0)$ (crosses) as functions of $L_{\rm
cube}$. Errors are estimated assuming Poisson statistics and that
there are $(L_{\rm GIF}/L_{\rm cube})^3$ independent volumes in the
simulation, where $L_{\rm GIF}=141.3h^{-1}$Mpc is the size of the GIF
$\Lambda$CDM simulation volume. This is known to underestimate the
true errors (e.g. \pcite{kim98}), but is sufficient for our present
purposes. Note that placing all galaxies at the halo centre (solid
squares), or placing one galaxy at the centre and making satellite
galaxies trace the dark matter of their halo (open squares) has little
effect on the measured $S(0)$, i.e. edge effects are unimportant for
this sample. For the smallest cells we consider $Q(0)$ accounts for
around 65\% of the value of $S(0)$, and makes a smaller contribution
for the larger cells. Also shown is the value of $S(0)$ predicted by
eqn.~(\ref{eq:maineq}) with the recovered value of $P_1(0)$ (dashed
line) and the true value of $P_1(0)$ (solid line). For the larger cell
sizes neither gives a good fit to the mock data points. This is due to
the failure of our assumption that $P_1(N;M)$ is roughly constant
throughout the mass bin (as discussed in \S\ref{sec:method}). However,
as we discuss below this does not drastically alter the recovered
values of $P_1(N)$.

\begin{figure}
\psfig{file=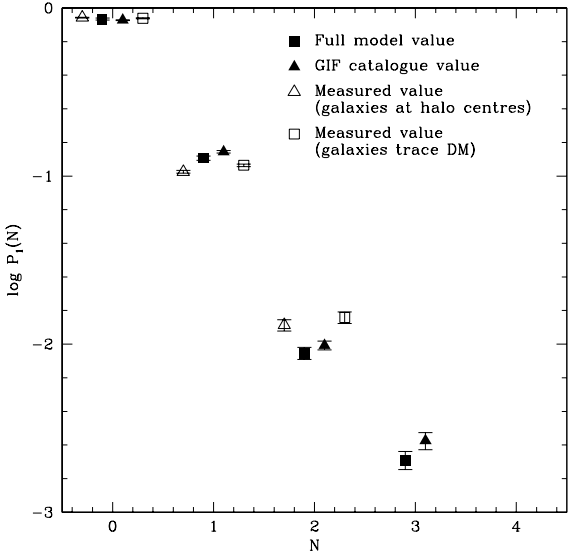,width=80mm}
\caption{$P_1(N)$ for halos more massive than $4\times
10^{11}h^{-1}$Mpc and galaxies brighter than $M_{\rm K}-5\log
h=-23.5$. Solid squares show $P_1(N)$ taken directly from the full
model of \protect\scite{benson00}, while solid triangles show that
taken directly from the GIF synthetic galaxy catalogue of
\protect\scite{benson00}. Open triangles show the $P_1(N)$ recovered
from the GIF synthetic galaxy catalogue via determinations of $S(N)$
when all galaxies are placed at the centre of their halo (error bars
are computed assuming Poisson statistics), while open squares show
the result when satellite galaxies are made to trace the dark matter
content of their halo (errorbars computed from $\Delta \chi^2$). The
points are offset slightly in $N$ for clarity.}
\label{fig:mockres}
\end{figure}

We determine $P_1(N)$ from the measured $S(N)$ and $Q(N)$ by solving
eqn.~(\ref{eq:maineq}) for $P_1(N)$ by minimizing $\chi^2$ (as
described in \S\ref{sec:method}). Figure~\ref{fig:mockres} shows the
true $P_1(N)$ as measured directly from the full model and from the
GIF synthetic galaxy catalogue (solid squares and solid triangles
respectively), with errorbars computed assuming Poisson statistics,
and the $P_1(N)$ recovered from the synthetic galaxy catalogue via the
$S(N)$'s with all galaxies at their halo centre (open triangles) and
with satellite galaxies tracing the dark matter of their halo (open
squares), with errorbars estimated from $\Delta \chi^2$. The first
three $P_1(N)$ are recovered with reasonably accuracy from the
synthetic galaxy catalogues. (For $N=0$, 1 and 2 the recovered
$P_1(N)$ differ from the true values by 3\%, 17\% and 46\%
respectively, although we caution that these values are from a single
realization of the synthetic galaxy catalogue and so may not be
representative.)

A weakness of this approach is that the equation for $P_i(N)$ depends
upon all $P_i(N^\prime)$ where $N^\prime<N$. Hence, any errors in the
determination of $P_i(0)$ affect the estimate of $P_i(1)$ etc. In the
case of the synthetic galaxy catalogues used here we can recover
$P_1(N)$ accurately for $N=0,1,2$. When we consider $P_1(3)$, however,
we find that the contribution to $S(3)$ from terms involving only
$P_1(0)$, $P_1(1)$ and $P_1(2)$ already exceeds the measured
value. Thus the solution to the equation requires that $P_1(3)$ be
negative, which is of course impossible. Thus with a dataset of this
size only the first few $P_1(N)$ can be measured.

\section{Discussion}
\label{sec:discuss}

We have described how the distribution of galaxies amongst halos, as
described by the function $P(N;M)$ (the probability of finding $N$
galaxies brighter than a specified luminosity $L_0$ in a halo of mass
$M$), can be measured directly from a galaxy redshift survey once a
model for the spatial distribution of dark matter halos is
assumed. Specifically we derive relations between the observationally
measurable quantity $S(N)$ (the probability of finding $N$ galaxies in
a cell) and the theoretically determinable quantity $Q(N_1,N_2,\ldots
,N_n)$ (the probability of finding different numbers of dark-matter
halos in a cell). These relations depend upon $P(N;M)$, thereby
allowing $P(N;M)$ to be determined from observational determinations
of $S(N)$ and a model of structure formation.

The distribution function $P(N;M)$ provides a complete description of
galaxy bias (at least on scales larger than the size of halos) in
terms of physically meaningful quantities and will also be sensitive
to the merging history of dark matter halos and the rate of
galaxy-galaxy mergers within dark matter halos. We have presented the
technique in its simplest form. We defer a more detailed study of
errors (including edge effects) and the limitations imposed by the
simplifying assumptions made to a future paper.

Our approach assumes a model for the underlying distribution of dark
matter halos, and the results obtained will therefore be dependent on
that model. Measurements of key cosmological parameters, perhaps from
measurements of the cosmic microwave background
\cite{jungman96,bond97}, and the dark matter power spectrum, from weak
lensing (e.g. \pcite{tyson00}) or Lyman-$\alpha$ forest studies
\cite{croft98}, in the near future should allow the halo distribution
to be fully determined.

Using the mock galaxy catalogues produced by \scite{benson00} we have
shown that $P(N;M)$ averaged over all halos more massive than $4\times
10^{11}h^{-1}M_\odot$ is measurable for the first few values of $N$
from a combination of the 2MASS dataset with a redshift survey such as
the SDSS or 2dFGRS. To measure $P(N;M)$ for higher $N$ or as a
function of $M$ would require larger datasets and a detailed
consideration of edge effects. Measurement of this quantity from
forthcoming galaxy redshift surveys will therefore provide strong
constraints for models of galaxy formation and clustering and reveal a
great deal about the connection between galaxies and dark matter.

\section*{Acknowledgments}

We would like to thank Marc Kamionkowski for a careful reading of this
work, Carlton Baugh, Shaun Cole, Carlos Frenk and Cedric Lacey for
making available results from their galaxy formation model, Simon
White for useful discussions and the Virgo Consortium for making
available the GIF simulations used in this work.

\appendix

\section{The Two-Point Correlation Function of Galaxies}
\label{ap:xi}

The two-point correlation function is a familiar clustering statistics
easily expressed in terms of $P(N;M)$. Suppose there is a halo of mass
$M_1$ to $M_1+\d M_1$ in a small volume element $\d V_1$. Let $\d
Q_{12}$ be the probability of finding a halo of mass $M_2$ to $M_2+\d
M_2$ in a small volume element $\d V_2$ a distance $r$ away from the
first halo. We can write
\begin{equation}
\d Q_{12}(r) = (1+\xi_{12}(r)) {\d n \over \d M}(M_1) {\d n \over \d M}(M_2) \d M_1 \d M_2 \d V_1 \d V_2,
\end{equation}
where $\xi_{12}(r)$ is the cross correlation function of these
halos. A single halo pair may contribute many galaxy pairs. On average
the above halo pair will contribute
\begin{equation}
\d N_{12}(r) = \bar{N}(M_1) \bar{N}(M_2) \d Q_{12}(r)
\label{eq:n12clus}
\end{equation}
galaxy pairs, where $\bar{N}(M)=\sum_{i=0}^\infty i P(i;M)$ is the
mean number of galaxies in a halo of mass $M$. For a random
distribution of galaxies we would expect
\begin{equation}
\d N^{(\rm r)}_{12}(r) = \bar{N}(M_1) \bar{N}(M_2) {\d n \over \d M}(M_1) {\d n \over \d M}(M_2) \d M_1 \d M_2 \d V_1 \d V_2.
\label{eq:n12ran}
\end{equation}
Integrating eqns.~(\ref{eq:n12clus}) and (\ref{eq:n12ran}) over all
halo masses we find the total number of galaxy pairs in the clustered
and random cases to be
\begin{eqnarray}
\d N_{gg}(r) & = & \int_{M_0}^\infty \int_{M_0}^\infty \bar{N}(M_1) \bar{N}(M_2) (1+\xi_{12}(r)) {\d n \over \d M}(M_1) {\d n \over \d M}(M_2) \d M_1 \d M_2 \d V_1 \d V_2 \\
\d N^{(\rm r)}_{gg}(r) & = & \int_{M_0}^\infty \int_{M_0}^\infty \bar{N}(M_1) \bar{N}(M_2) {\d n \over \d M}(M_1) {\d n \over \d M}(M_2) \d M_1 \d M_2 \d V_1 \d V_2 = n_{\rm gal}^2 \d V_1 \d V_2,
\end{eqnarray}
where $n_{\rm gal}$ is the mean number density of the galaxies. The
galaxy-galaxy correlation function is defined to be
\begin{eqnarray}
\xi_{\rm gg}(r) & = & {\d N_{gg}(r) \over \d N^{(\rm r)}_{gg}(r)} - 1 \nonumber \\
 & = & \int_{M_0}^\infty \int_{M_0}^\infty \xi_{12}(r) {\bar{N}(M_1) \bar{N}(M_2) \over n_{\rm gal}^2} {\d n \over \d M}(M_1) {\d n \over \d M}(M_2) \d M_1 \d M_2.
\end{eqnarray}


\begin{thebibliography}{}
\bibitem[Baugh et al. <1998>]{baugh98}Baugh~C.~M., Cole~S., Frenk~C.~S., Lacey~C.~G., 1998, ApJ, 498, 504
\bibitem[Benson et al. <2000a,b>]{ajbdummy}
\bibitem[Benson et al. <2000a>]{benson00}Benson~A.~J., Cole~S., Frenk~C.~S., Baugh~C.~M., Lacey~C.~G., 2000a, MNRAS, 311, 793
\bibitem[Benson et al. <2000b>]{benson00b}Benson~A.~J., Baugh~C.~M., Cole~S., Frenk~C.~S., Lacey~C.~G., 2000b, MNRAS, 316, 107
\bibitem[Bond, Efstathiou \& Tegmark <1997>]{bond97}Bind~J.~R., Efstathiou~G., Tegmark~M., 1997, MNRAS, 291, 33
\bibitem[Blanton et al. <2000>]{blanton00}Blanton~M. et al., 2000, AAS, 196, \#53.12
\bibitem[Brinchmann \& Ellis <2000>]{jarle00}Brinchmann~J., Ellis~R.~S., 2000, ApJL in press (astro-ph/0005120)
\bibitem[Bullock et al. <2000>]{bullock99}Bullock~J.~S., Kolatt~T.~S., Sigad~Y., Somerville~R.~S., Kravtsov~A.~V., Klypin~A.~A., Primack~J.~R., Dekel~A., MNRAS in press (astro-ph/9908159)
\bibitem[Cole et al. <2000>]{cole00}Cole~S.~M., Lacey~C.~G., Baugh~C.~M., Frenk~C.~S., 2000, MNRAS in press (astro-ph/0007281)
\bibitem[Croft et al. <1998>]{croft98}Croft~R.~A.~C., Weinberg~D.~H., Katz~N., Hernquist~L., 1998, ApJ, 495, 44
\bibitem[Dalton <2000>]{dalton00}Dalton~G.~B. et al., 2000, AAS, 196, \#56.05
\bibitem[Diaferio et al. <1999>]{diaferio99}Diaferio~A., Kauffmann~G., Colberg~J.~M., White~S.~D.~M., 1999, MNRAS, 307, 537
\bibitem[Governato et al. <1998>]{fabio98}Governato~F., Baugh~C.~M., Frenk~C.~S., Cole~S., Lacey~C.~G., Quinn~T., Stadel~J., 1998, Nat., 392, 359
\bibitem[Jungman et al. <1996>]{jungman96}Jungman~G., Kamionkowski~M., Kosowsky~A., Spergel~D.~N., 1996, Phys. Rev. D, 54, 1332
\bibitem[Kauffmann, Nusser \& Steinmetz <1997>]{kns}Kauffmann~G., Nusser~A., Steinmetz~M., 1997, MNRAS, 286, 795
\bibitem[Kauffmann \& Charlot <1998>]{kc98}Kauffmann~G., Charlot~S., 1998, MNRAS, 297, L23
\bibitem[Kauffmann et al. <1999a,b>]{kauffdummy}
\bibitem[Kauffmann et al. <1999a>]{kauff99a}Kauffmann G., Colberg J.~M., Diaferio A., White S.~D.~M., 1999a, MNRAS, 303, 188
\bibitem[Kauffmann et al. <1999b>]{kauff99b}Kauffmann G., Colberg J.~M., Diaferio A., White S.~D.~M., 1999b, MNRAS, 307, 529
\bibitem[Kim \& Strauss <1998>]{kim98}Kim~R.~S.~J., Strauss~M.~A., 1998, ApJ, 493, 39
\bibitem[Lemson \& Kauffmann <1999>]{lemson99}Lemson~G., Kauffmann~G., 1999, MNRAS, 302, 111
\bibitem[Peacock \& Smith <2000>]{peacock00}Peacock~J.~A., Smith~R.~E., 2000, MNRAS, 318, 1144
\bibitem[Press \& Schechter <1974>]{PS74}Press~W.~H., Schechter~P., 1974, ApJ, 187, 425
\bibitem[Press et al. <1992>]{numrec}Press~W.~H., Flannery~B.~P., Teukolsky~S.~A., Vetterling~W.~H., 1992, Numerical Recipes: The Art of Scientific Computing (2nd ed.; Cambridge: Cambridge Univ. Press)
\bibitem[Seljak <2000>]{seljak00}Seljak~U., 2000, MNRAS, 318, 203
\bibitem[Skrutskie et al. <1995>]{skrutskie95}Skrutskie~M.~F. et al., 1995, AAS, 187, \#75.07
\bibitem[Somerville et al. <2000>]{somerville00}Somerville~R.~S., Lemson~G., Sigad~Y., Dekel~A., Kauffmann~G., White~S.~D.~M., submitted to MNRAS (astro-ph/9912073)
\bibitem[Tyson, Wittman \& Angel <2000>]{tyson00}Tyson~J.~A., Wittman~D., Angel~J.~R.~P., 2000, in ``Dark Matter 2000: 4th International Symposium on Sources and Detection of Dark Matter/Energy in the Universe'', Springer (astro-ph/0005381)
\bibitem[Vogt et al. <1997>]{vogt97}Vogt~N.~P., Phillips~A.~C., Faber~S.~M., Gallego~J., Gronwall~C., Guzman~R., Illingworth~G.~D., Koo~D.~C., Lownethal~J.~D., 1997, ApJ, 479, 121
\bibitem[Wechsler et al. <2000>]{wechsler00}Wechsler~R.~H., Somerville~R.~S., Bullock~J.~S., Kolatt~T.~S., Primack~J.~R., Blumenthal~G.~R., Dekel~A., 2000, submitted to ApJ (astro-ph/0011261)
\end{thebibliography}
\end{document}